\documentclass[12pt,preprint]{aastex}

\slugcomment{ }

\shorttitle{MS\,2053.7-0449}
\shortauthors{Verdugo, T., de Diego & Limousin}

\begin{document}

\title{MS\,2053.7-0449: Confirmation of a bimodal mass
distribution from strong gravitational lensing}

\author{T. Verdugo and J.A. de Diego}
\affil{Instituto de Astronom\'{\i}a, UNAM, AP 70-264, 04510 Mexico
DF} \email{tomasv, jdo@astroscu.unam.mx}

\and

\author{Marceau Limousin}
\affil{Dark Cosmology Centre, Niels Bohr Institute, University of Copenhagen,
Juliane Marie Vej 30, 2100 Copenhagen, Denmark}
\email{marceau@dark-cosmology.dk}

\begin{abstract}

We present the first strong lensing study of the mass
distribution in the cluster MS\,2053-04 based on HST archive
data. This massive, X-ray luminous cluster has a redshift
z=0.583, and it is composed of two structures that are
gravitationally bound to each other. The cluster has one
multiply imaged system constituted by a double gravitational
arc.

We have performed a parametric strong lensing mass
reconstruction using NFW density profiles to model the cluster
potential. We also included perturbations from 23 galaxies,
modeled like elliptical singular isothermal sphere, that are
approximately within $1'\times1'$ around the cluster center.
These galaxies were constrained in both the geometric and
dynamical parameters with observational data. Our analysis
predicts a third image which is slightly demagnified. We found
a candidate for this counter-image near the expected position
and with the same F702W-F814W colors as the gravitational arcs
in the cluster. The results from the strong lensing model shows
the complex structure in this cluster, the asymmetry and the
elongation in the mass distribution, and are consistent with
previous spectrophotometric results that indicate that the
cluster has a bimodal mass distribution. Finally, the derived
mass profile was used to estimate the mass within the arcs and
for comparison with X-ray estimates.

\end{abstract}


\section{INTRODUCTION}\label{introd}

The possibility that clusters of galaxies might produce
gravitational lensing had been considered by several authors
\citep {kar76,dyr76, nar84}. Twenty years ago \citep {lyn86}
and \citep {sou87} independently announced the detection of an
arc-like structure around the center of Abell 370, but it was
\citet {pac87} who explained the phenomenon in terms of
gravitational lensing. Currently, the strong and weak regimes
of the gravitational lensing effect are often used for
constraining the density profiles of galaxy clusters.

The massive clusters of galaxies have a surface density larger
than the critical value for strong lensing, approximately 1.0 g
cm$^{-2}$ \citep {tur84}. They can produce multiple, strongly
elongated images (giants arcs) of background galaxies lying on
top of the caustic lines generated by the cluster potential.
Observationally, the first dedicated search for gravitational
arcs in X-ray clusters was performed by  \citet {lef94} using
the \emph {Einstein Observatory} Extended Medium Sensitivity
Survey (EMSS). A subsequent search was performed by \citet
{lup99}, which allowed to confirm the relationship between
X-ray luminosities and cluster masses; these authors found
strong lensing in eight of 38 clusters. A subsequent survey
based on optically selected clusters in the Las Campanas
Distant Cluster Survey confirmed these giant-arc fractions
\citep {zar03} and \citet{gla03} reports even larger fractions
in the Red-Sequence Cluster Survey, showing that giant arcs are
quite common in clusters of galaxies.

Since the shapes of these arcs are related to the gravitational
potential, the strong lensing offers a method to investigate
how the barionic and dark matters are distributed in the inner
regions of the cluster. Also it is possible to put significant
constraints on the cosmological parameters using a set of
strong lensing clusters showing multiple images with
spectroscopic redshifts  \citep{sou04}. Mass distributions have
been modeled for many clusters with giants arcs. In a few
cases, accurate models for the mass distribution have led to
the identification of additional counter-image candidates
\citep {kne96}. In other cases, the detection of the predicted
images is uncertain, since they are lost in the cD light
distribution \citep[see][]{gav03,san04}. Recently, the
gravitational lens cluster A\,1689 has been modeled in
unprecedented detail by \citet{bro05,lim06}. These authors
combined HST Deep Advanced Camera images and extensive ground
based spectroscopy, identifying 32 multiply lensed systems.

The aim of this work is to describe the density profile and the
mass distribution of MS\,2053, using strong lensing. This
cluster has a redshift of z=0.583, is not very optically
rich in comparison with other z $>$  0.5 EMSS clusters
\citep{lup99}, and among these
high-redshift clusters it has the lowest X-ray luminosity.
BeppoSAX observations yield a X-ray luminosity $L_X$(2-10 keV)
= (3.9$\pm$1.0)$\times10^{44}$ erg s$^{-1}$\citep{del00}, and
Chandra $L_X$(0.5-2 keV) = 3.5 $\times10^{44}$ erg s$^{-1}$
\citep{vik02}. \citet{lup92} discovered a gravitationally
lensed blue arc in deep images of MS\,2053, located $\sim 16
\arcsec$ in the North direction from the Brightest Cluster
Galaxy (BCG). The arc is $\sim 11 \arcsec$ long and breaks into
two clumps, labeled A and B. Recently, \citet{tra05} were able
to measure the redshift of image A using Keck spectroscopy data
and determining the source redshift to be z = 3.146.

\citet{hoe02} detected the weak lensing (WL) signal induced by
this cluster of galaxies from a two-colour mosaic of six Hubble
Space Telescope (HST) Wide Field and Planetary Camera 2 (WFPC2)
images. They fitted a singular isothermal sphere model to the
observed azimuthally averaged tangential distorsion and
estimated the cluster velocity dispersion $\sigma = 886$ km
s$^{-1}$. \citet{hoe02} also fitted the predicted profile from
the NFW halo to the observed tangential distorsion of MS\,2053,
and found a concentration parameter (see section 3.1) in good
agreement with the predicted value for a cold dark matter model
without cosmological constant. \citet{tra05}  combined images
from the wide-field HST WFPC2 and extensive spectral data from
the Keck  Low Resolution Imager Spectrograph (LRIS) to make a
detailed study of the galaxy populations in MS\,2053. They
found that the cluster is composed of a main cluster (MS
2053-A) and an infalling structure (MS\,2053-B) that are
gravitationally bound to each other; their study also shows
that the extended spatial distribution and lack of spectral
segregation in MS\,2053-B is consistent with a scenario in
which it is in the initial stages of being accreted by the main
cluster. Of the 149 spectroscopically confirmed cluster
members, 113 belong to MS\,2053-A and 36 to MS\,2053-B. From
these previous researches we present in Table~\ref{tbl-1} a
compilation of physical quantities relevant to our work .

Despite all the research cited above, only \citet{hoe02} have
made an attempt to describe the strong lensing effect for this
cluster and it was done when the arc redshift had  not been
measured yet. Assuming a redshift of z=2 for the arc and
adopting a singular isothermal sphere, these authors found a
velocity dispersion of about 1030 km s$^{-1}$, consistent with
their WL estimate. They argue that the mass distribution is
elongated in the direction of the arc, which can be very well
appreciated in the light distribution, and this elongation is
the cause of the high value of their strong lensing mass
estimate. Such elongation in the distribution of the cluster
members was also reported by \citet{tra05}. To assess with
precision this mass distribution, we present a detailed strong
lensing model of the bimodal cluster MS\,2053.

In $\S$\,\ref{observ}, we describe the observational data and
present our sample of field galaxies.  In $\S$\,\ref{model}, we
define the model of the lensing cluster, and depict the
profiles for both the cluster and the Galaxy-scale mass
components. In $\S$\,\ref{results}, we present the main
results. In $\S$\,\ref{discuss}, we summarize and discuss our
results. Finally, in $\S$\,\ref{conclus}, we present the
conclusions. Throughout this paper, we adopt a spatially-flat
cosmological model dominated by cold dark matter and a
cosmological constant. We use $\Omega_{m}=0.3$,
$\Omega_{\Lambda}=0.7$ and H$_0$ = 70 km s$^{-1}$ Mpc$^{-1}$.
With these cosmological parameters and at the redshift of
MS\,2053, $1 \arcsec$ corresponds to 4.62 $h^{-1}$ kpc.

\section{THE OBSERVATIONAL DATA}\label{observ}

\begin{figure}
 \epsscale{.80}
 \plotone{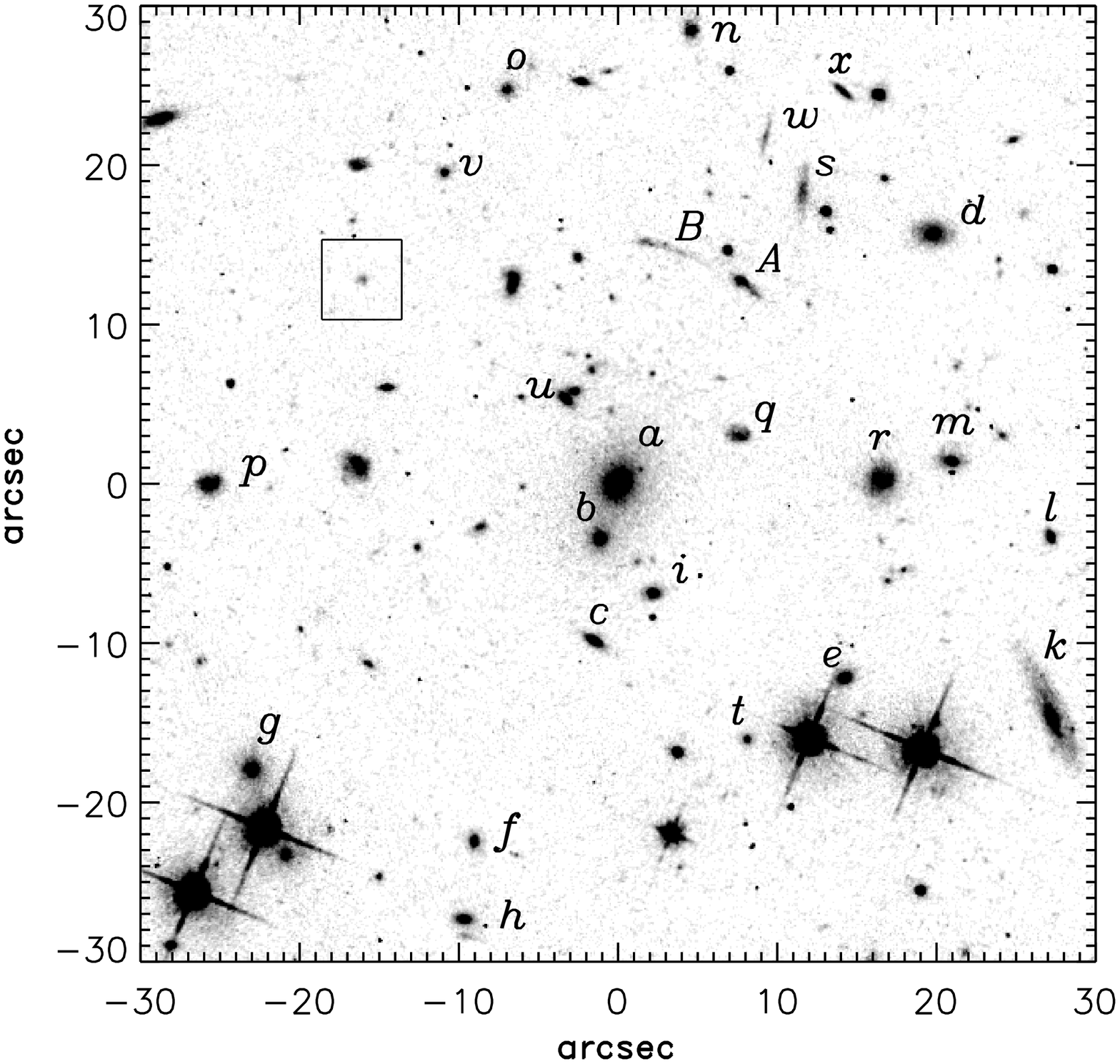}
 \caption{HST WFPC2 F814W image of the central region ($60\arcsec
\times60\arcsec$) of the MS2053 cluster, North is up, East is left.
The arcs A and B are identified, as well as the 23 galaxies
considered in our models (lower-case letters). The square
shows the region enlarged in Fig. \ref{fig3}. \label{fig1}}
\end{figure}

The HST data have been obtained from the Multimission Archive
at Space Telescope (MAST). They consist of two WFPC2
images\footnote{Proposal ID 5991} obtained in both the F814W
and the F702W filters; the total integration time for each
filter was 2600 and 2400~s, respectively. The image reduction
was performed using the IRAF/STSDASS package. First, a
warm-pixel rejection was applied to the images using the IRAF
task \emph{warmpix}. The cleaned images were then combined with
the task \emph{crrej} to remove cosmic-rays hits. Finally,  the
background was subtracted and the WFPC2 chips were combined
using the task \emph{wmosaic}. In Figure~\ref{fig1} we show a
$60''\times60''$ field obtained from the final four-chip mosaic
frame for the F814W filter; this is the region used in our
analysis of strong lensing.

The magnitudes and geometric parameters of the galaxies were
measured using SExtractor \citep{ber96}. For the arc photometry
we used a different procedure because SExtractor often
overestimate the sky background \citep[see][]{smi05}, and the
arcs are highly distorted. These two circumstances reduce the
precision of automatic photometric measurements. Therefore, we
employed polygonal apertures to obtain more accurate
measurements. The vertices of the polygons for each arc were
determined using the IRAF task \emph{polymark}, and the
magnitudes inside this apertures were calculated using the IRAF
task \emph{polyphot}. Hence, we found F814W magnitudes of
m$_{F814W}=21.64\pm0.02$ and m$_{F814W}=22.01\pm0.04$ for arcs
A and B respectively, and therefore the total magnitude for the
system is $21.06\pm0.11$. Similarly, the F702W magnitudes are
m$_{F702W}=22.05\pm0.02$ for arc A and m$_{F702W}=22.38\pm0.04$
for arc B, contributing to a total magnitude of $21.45\pm0.11$.
Our measurements of the total brightness in these two bands are
slightly lower than the values reported by \citet{san05}
(m$_{F814W}=20.90$ and m$_{F702W}=21.27$). The small
discrepancies probably arise from the different methodologies
used to calculate the total magnitudes. For example,
\citet{san05} used a single aperture to measure the magnitude
for the system of arcs as a whole (arc AB), while we used two
apertures to measure each arc independently.

We also calculated the F702W-F814W color for the AB arc system.
The color for arcs A ($0.41 \pm 0.03$) and B ($0.37 \pm 0.06$),
are similar within the errors, and the color for the whole AB
system, $0.39 \pm 0.15$, is similar to the value of 0.37
reported by \citet{san05}.

The general properties of the cluster members are presented in
Table~\ref{tbl-2}. We used the relative positions for the
individual galaxies provided by \citet{tra02} to calculate the
coordinates of each member of the cluster. Column (1) lists the
identification for each MS\,2053 galaxy. Columns (2) and (3)
list the right ascension and declination. Column (4), the F814W
magnitudes. Columns (5) and (6) show the geometric parameters
derived from SExtractor: ellipticity $\epsilon$ and the
position angle $\theta_0$,  which gives the orientation of the
semi-major-axis from the horizontal line in the image, measured
counter-clockwise. Column (7) lists the redshifts. Finally,
columns (8) and (9) list the values and the references for the
central velocity dispersions. The upper limits were taken from
\citet{tra05} (see their Figure 11), who consider only galaxies
brighter than M$_{Be}$ $\leq -20 + 5\log h$.

\section{THE MODEL OF THE LENS}\label{model}

\subsection{Dark matter component}

For studying the mass distribution in MS\,2053, we use standard
lens modeling techniques implemented in the LENSTOOL\footnote{
This software is publicly available at:
http://www.oamp.fr/cosmology/ lenstool/} ray-tracing code
\citep{kne93}.

We model the cluster with a NFW profile that has been predicted in
cosmological N-body simulations \citep{nav97}:

\begin{equation}\label{eq:rho}
\rho(r) = \frac{\rho_{crit}(z)\delta_{c}(z)}{(r/r_s)(1+r/r_s)^{2}}
\end{equation}

\noindent where $\delta_c$ is a characteristic density
contrast, $\rho_{crit}(z)$ is the critical density of the
universe as a function of the redshift $z$, and $r_s$ is a
scale radius that corresponds to the region where the
logarithmic slope of the density equals the isothermal value.
However, higher-resolution simulations have suggested that the
inner cusp of the NFW profiles is too shallow
\citep[e.g.][]{moo98, ghi00}, giving origin to the generalized
NFW-type profile. It is worth noting that \citet{gav05} studied
MS 2137 - 2353, obtaining similar results by using both
profiles for the strong lensing regime. Therefore, for
simplicity we adopt a profile similar to the original NFW.
However, we consider an elliptical profile since in
modeling the gravitational lens it is important to
take into account the asphericity in the cluster potential.

To generalize the spherical model to an ellipsoidal one, we
used the "pseudo-elliptical" NFW proposed by \citet{gol02}.
This potential is characterized by six parameters: the position
$X,Y$; the ellipticity $\epsilon$; the position angle
$\theta$; $r_s$ and the velocity dispersion  $\sigma_s$. We
define $\sigma_s$ as:

\begin{equation}\label{eq:sigma}
\sigma_s^{2} = 4\left(1+\ln{\frac{1}{2}}\right)G r_s^{2} \rho_{crit}\delta_{c}
\end{equation}

\noindent This is approximately half of the value proposed by
\citet{gol02} and it is defined in such a way that it
represents a realistic velocity dispersion at radius  $r_s$ and
not only a scaling parameter (see the appendix).

We can relate this $\sigma_s$ to the virial mass $M_{\Delta}$
and the virial radius $r_{\Delta}$ of the cluster. If we define
$r_{\Delta}$ as the radius of a spherical volume within which
the mean density is $\Delta$ times the critical density at the
given redshift $z$ ($M_{\Delta} = \frac{4}{3} \pi
r_{\Delta}^{3} \rho_{crit}\Delta$), then the virial radius can
be expressed by:

\begin{equation}\label{eq:rdelta}
r_{\Delta} =\left[ 2M_{\Delta}G/\Delta H(z)^{2} \right]^{1/3}
\end{equation}

\noindent  The value of the characteristic overdensity $\Delta$
is taken from the solution to the collapse of a spherical
top-hat model. Its value depends on $\Omega$ and can be
approximated by $ \Delta  = 178\Omega_{m}^{0.45}$ when
$\Omega_{m} + \Omega_{\Lambda} = 1$ \citep{lac93,eke96,eke98}.
Integrating equation \ref{eq:rho} and using the above
definition, it is straightforward to show that the ratio
between virial and scale radii, which is  commonly called the
concentration, $c = r_{\Delta}/r_s$, is related to $\delta_c$
by

\begin{equation}\label{eq:delta}
\delta_{c} = \frac{\Delta}{3}
\frac{c^3}{ \left[ \ln(1+c)-\frac{c}{1+c} \right] }
\end{equation}

\noindent Solving equation \ref{eq:delta}, we can determine the
concentration value given the $z$, $\sigma_s$ and $r_s$.

\subsection{Galaxy-scale mass components}

For simplicity, we modeled the 23 galaxies using elliptical
singular isothermal spheres (ESISs) for the individual galaxies
\citep{bla87}. This galaxy-scale mass components were included
as perturbation into the cluster potential. The
potential for ESISs is given by\\

\begin{equation}\label{eq:varphi}
\varphi(r,\theta) = 4\pi \frac{ \sigma_0^{2} }{c ^{2} }\frac{ D_{ LS
} }{ D_{ OS } } r \sqrt{ 1+ \epsilon\cos(2(\theta - \theta_{0})) }
\end{equation}

\noindent where $\sigma_0^{2}$ corresponds exactly to the true
3D velocity dispersion and $\theta_{0}$ is the position angle.
Note that equation \ref{eq:varphi} is defined completely if we
can obtain the values for  $\sigma_0^{2}$, the ellipticity
$\epsilon$, and $\theta_{0}$ from observational data. The
luminosity distribution of a given galaxy does not trace the
dark mass distribution in its halo, however there is evidence
that the projected mass and light distributions tend to be
aligned \citep{kee98}. \citet{koo06} report this alignment
between light and mass for a sample of early-type lens
galaxies, for which the isophotal and isodensity countour trace
each other. These authors also report an alignment between the
stellar component and the singular isothermal ellipsoids used
in their lens models. Following these authors, we infer the
parameters of the galaxy-scale mass components from the
position, the ellipticity and the position angle of the
luminous component.

\citet{wuy04} have measured the velocity dispersions of several
elliptical and lenticular galaxies in the cluster, and
\citet{tra05} have estimated the internal velocity dispersion
for the whole cluster sample with M$_{Be}$ $\leq -18 + 5\log h$.
Nevertheless, the individual data for each galaxy has not been
published for all the objects. To minimize the number of model
free parameters, the galaxies for which the velocity dispersion
is not available (see Table~\ref{tbl-2}) are scaled as a
function of luminosity following the Faber-Jackson relation
\citep{fab76}:

\begin{equation}\label{eq:k}
\sigma_0 = \sigma_0^{*}(\frac{L}{L^{*}})^{1/4}
\end{equation}

\noindent which transformed to magnitudes is expressed by
$\log\sigma_0=-0.1m + K$, where $\emph{K}$ is a scaling factor.
In this way the cluster members are incorporated into the model
as a galaxy scale perturbation with all their individual
parameters fixed.

\subsection{The model optimization}

In principle, the optimizations in the source plane and in the
image plane are mathematically equivalent. However, we chose
using the source plane because the solutions are numerically
simpler and faster to compute. The fit of the model is
optimized by mapping the positions of the multiple images back
to the source plane and requiring them to have a minimal
scatter. To achieve this goal we use a
figure-of-merit-function, $\chi^{2}$, to quantify the goodness
of the fit for each trial of the lens model i.e., a function
that measures the agreement between our data and the fitting
model for a particular choice of the parameters:

\begin{equation}\label{eq:chi2}
\chi^{2} = \chi_{pos}^{2}+\chi_{shape}^{2}
\end{equation}

\noindent The first term is constructed as follows: given a
model, we compute the position in the source plane $\vec
u_i^{S}=(x_i^{S},y_i^{S})$ for each observed image ($1 \leq i
\leq N$), using the lens equation, $ \vec u_i^{S} = \vec
u_i^{I} - \nabla\varphi( \vec u_i^{I})$, where $\varphi$ is the
lens potential. Therefore, the barycenter $\vec u^{B} =
(x^B,y^B)$ is constructed from the N sources, and the
$\chi^{2}$  for the position is defined by:

\begin{equation}
\chi_{pos}^{2} =     \frac{1}{N}\sum_{i=1}^{N-1}
\frac{\left(x^B-x_{i}^S\right)^{2}+\left(y^B-y_{i}^S\right)^{2}}{\sigma_{pos}^{2}}
\end{equation}

\noindent were $\sigma_{pos}$ is the error in determining the
position of the source.

The second term, $\chi_{shape}^{2}$, measures how well the
shape of the arcs is reproduced by the model. Given a model,
we compute the complex deformation, $\bar{\tau}_i$, for
each observed image ($1 \leq i \leq N$),using \emph{the lens
equation for complex deformation} \citep[see the discussion in]
[and their detailed derivation]{kne96}. The complex deformation
of the image represents how much the intrinsic source shape
changes due to the induced deformation caused by the potential.
This complex deformation is function of the ellipticity, the
position angle and the lens potential, $\bar{\tau}_i^{I} =
\bar{\tau}_i^{I}( \epsilon_i^{S},\theta_i^{S},\varphi)$.
It has an equivalent form in the source plane,
$\bar{\tau}_i^{S} =
\bar{\tau}_i^{S}( \epsilon_i^{I},\theta_i^{I},\varphi)$. If we
define $\tau_{x,i}^{S}$ and $\tau_{y,i}^{S}$ as the components
of the complex deformation in the source plane, and
$\sigma_{shape}$ as the error in determining the shape of the
source, we get:

\begin{equation}
\chi_{shape}^{2} =     \frac{1}{N}\sum_{i=1}^{N-1}
\frac{\left(\tau_{x,i}^{S}-\tau_{x,i-1}^{S}\right)^{2}+\left(\tau_{y,i}^{S}-\tau_{y,i-1}^{S}\right)^{2}}{\sigma_{shape}^{2}}
\end{equation}

\noindent The errors, $\sigma_{pos}$  and $\sigma_{shape}$,
have to be measured in the source plane, but these are not
directly observable quantities. To avoid this problem,we
assumed that the errors in the source plane arise from the
measurements performed in the image plane. Thus, we arbitrarily
set $\sigma_{pos}$  and $\sigma_{shape}$  to $0.2 \arcsec$ and
$0.8  \arcsec$ respectively, because these are the accuracies
achieved in the position and shape measurements from the HST
images. This is qualitatively correct in the sense that for
poor models, the merit function (equation \ref{eq:chi2}) gives
large values and thus indicates a poor fit.

In the 6.2 version  of LENSTOOL the Bayesian Markov Chain Monte
Carlo (MCMC) package BAYESIS\footnote{J. Skilling,
http://www.inference.phy.cam.ac.uk/bayesys/} has been
implemented. The Bayesian MCMC method will be described in a
dedicated forthcoming publication (Jullo et al. 2007, in prep.)
This package allows to obtain the errors and to avoid local
$\chi^{2}$ minima. According to a user defined model parameter
space, the MCMC sampler draws random models and compute their
$\chi^{2}$. Progressively, it converges to the most likely
parameter space and outputs the best solution. We used these
random models to perform the statistics to compute error bars
on the estimation of the parameters, and to constrain the
positions of the bimodal cluster (see next section).

\section{RESULTS}\label{results}

\subsection{A simple model for arc A}

The only arc for which the redshift has been measured is arc A
\citep{tra05}. The first step was to demonstrate that arcs A
and B belong to the same source. But a model that includes only
arc A and not arc B is unconstrained, and the data available is
not adequate to break the degeneracy. Given the general
characteristics of the cluster, and a gravitational lensed
source with the properties inferred from arc A (i.e., redshift
and arc position), we investigate if this source would produce
multiple images or not. To achieve this task, we constructed a
simple spherical model with a NFW profile and its central 23
galaxies modeled like elliptical isothermal spheres.

Using equation \ref{eq:delta} we calculated the virial radius
of the cluster. For a virial mass of $3.7
\times10^{15}$M$_\sun$ and a cluster redshift of z=0.583 (see
Table~\ref{tbl-1}), $r_{\Delta}$ $\sim 2.9 $ Mpc. Therefore,
$r_s$  $\sim 1.0 $ Mpc with a concentration parameter $c$=2.9
\citep[according to the prescription described in section 3.3
of ][]{eke01}. The characteristic velocity, $\sigma_s$ = 1600
km s$^{-1}$, is obtained from equation \ref{eq:sigma}. With
these values ($r_s$ and $\sigma_s$), the NFW profile is fully
characterized. At this stage, the ellipticity and the position
angle are set to zero.

\begin{figure}
 \epsscale{.80}
 \plotone{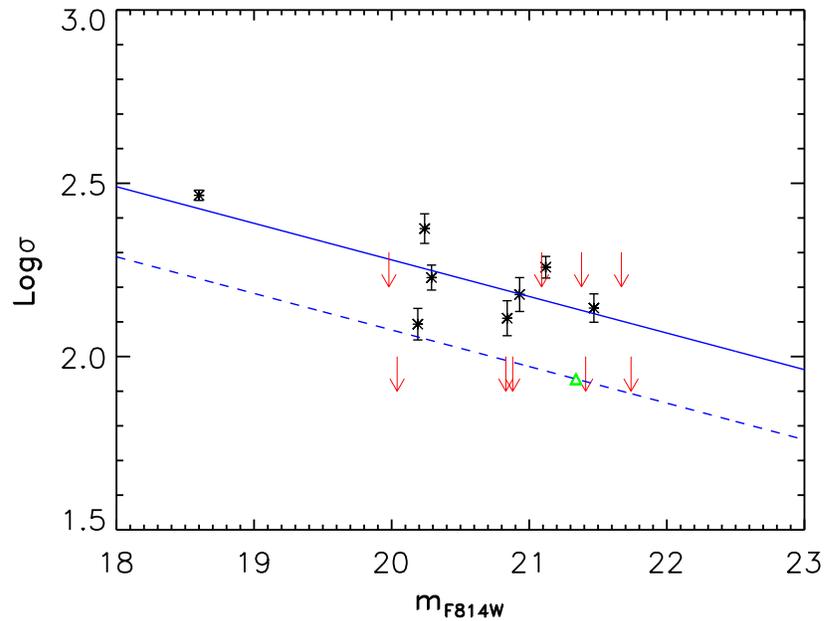}
 \caption{The cluster galaxies and the Faber-Jackson relation.
The solid line shows the correlation for the early-type galaxies
of \citet{wuy04} for which the velocity dispersion is available.
Dashed line is a similar correlation but for galaxies
not included in the \citet{wuy04} sample. The triangle
represents galaxy \emph{n}, whose velocity dispersion has
been measured by \citet{tra03}. \label{fig2}}
\end{figure}

The velocity dispersion in the galaxy-scale mass components was
fixed using the scaling relation given by equation \ref{eq:k}.
We found that the early-type cluster galaxies in the sample of
\citet{wuy04} satisfy the Faber-Jackson relation with
$\log\sigma_0=-(0.1\pm 0.05)m + (4.4\pm1.1)$. For the remaining
subset of galaxies not included in the \citet{wuy04} sample,
only galaxy $n$ has a measured $\sigma_0$ value. We supposed
that these galaxies also satisfy the relation but with a
different scaling factor $K$. In order to construct the new
correlation, we considered the luminosity $L^*$ corresponding
to the faintest galaxy in Table~\ref{tbl-2} (galaxy $w$,
m$_{F814W}=22.57$). We set $\sigma_0^* = 64.7$ km s$^{-1}$ such
that the galaxy $n$ satisfies this correlation. With these
values, the velocity dispersion in the subset of galaxies not
included in the \citet{wuy04} sample is expressed by
$\log\sigma_0=-0.1m + 4.19$. Figure~\ref{fig2} show both, the
previous $\log\sigma_0$ and the Faber-Jackson relations for the
early-type galaxies of \citet{wuy04}.

With all these parameters we calculate the counter-images of
arc A. The model obtained with these fixed parameters predicts
two more images, one very demagnified and the other only one
magnitude bellow the arc A; the position does not agree with B,
but this shows that for this cluster characteristics and with
this redshift of the source, A would have multiple
counter-images. Therefore, is entirely plausible that images A
and B arise from the same source.

\subsection{Bimodality evidence}

Two kinds of models are possible depending on whether we assume
that MS\,2053 is composed of one or two clusters, as reported
by \citet{tra05}. We performed the fits setting up the position
(two parameters), the velocity, scale radius, ellipticity, and
the position angle of the cluster as free parameters. These 6
parameters define the lens potential and they comprise all the
allowed free parameters for  this single cluster case. With
just 2 images and 4 constrains (if there is a total of
$\sum_{i=1}^{n}n_i=N$ images, then there are
$\sum_{i=1}^{n}4(n_i-1)=N_c$ constrains on the models assuming
that the position, ellipticity, and position angle of the
images are fitted) the model is underconstrained. And it gets
worse in the second model where we have 12 parameters defining
the lens potential. Therefore, we expect that there may be a
family of models that can fit the data, even with unrealistic
parameters. Trying to avoid this problem, we have limited one
parameter in the NFW, the scale radius $r_s$.

The $r_s$  cut-off was made in the following way:  Given the
cluster redshift and using equation \ref{eq:rdelta}, we can
write the scale radius as $r_s \propto M^{1/3}$ , being $M$ the
mass found from lensing. Since lensing is sensitive to the
integrated mass along the line of sight, we expect
discrepancies between the true 3D mass, $M_{true}$, and the
lensing mass estimates of our models, $M_{lens}$. Following
\citet{gav05}, we consider an axisymmetric NFW density profile
with either the major or the minor axis aligned toward the line
of sight. Therefore, the net effect of the projection is given
by $M_{lens} = qM_{true}$; where $q$ is the ratio between the
semiaxis aligned and the semiaxis perpendicular to the line of
sight. Thus, we can write:

\begin{equation}
r_s \propto q^{1/3}M_{true}^{1/3}
\end{equation}

\noindent Using the concentrations shown in Table~\ref{tbl-1}
and the axis values from \citet{gav05} we can constrain the
scale radius for each model. For the one component model, 0.60
Mpc $< r_s <$ 1.74 Mpc. Similarly for the two component model
0.30 Mpc $< r_s <$ 0.87 Mpc and 0.10 Mpc $< r_s <$ 0.28 Mpc for
component one and two respectively.\\

\begin{figure}
 \epsscale{.80}
 \plotone{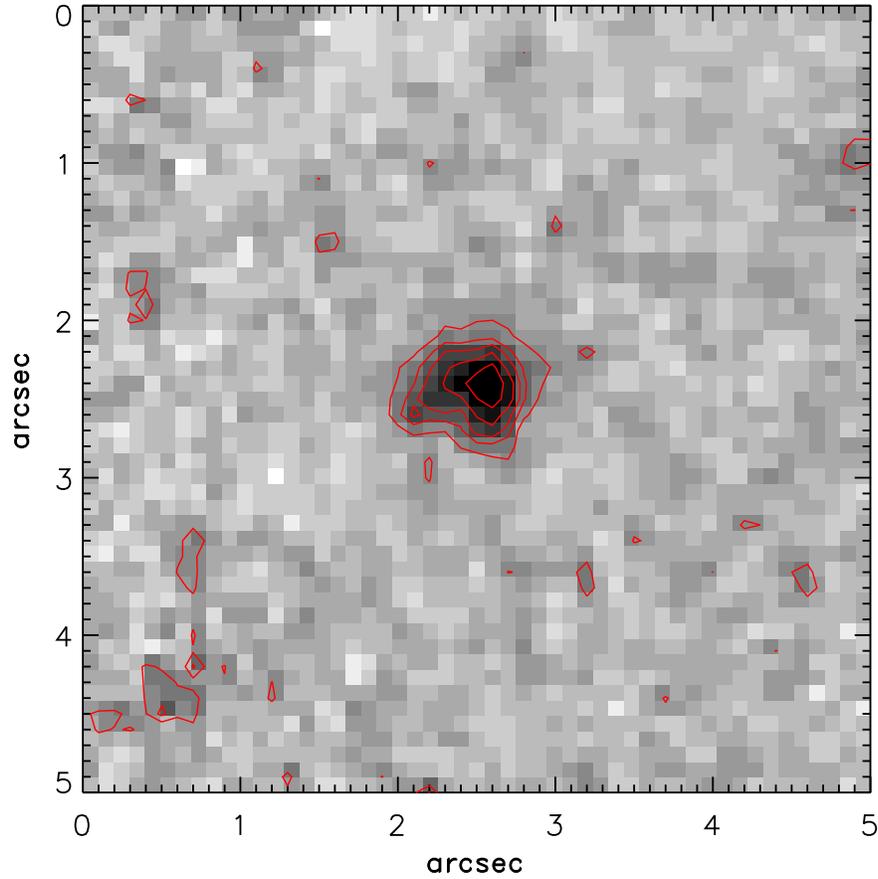}
 \caption{A detail of the HST observation in the F814W filter of
 the region shown in Fig. \ref{fig1}, centered on the
 counter-image candidate (North is up, East is left).
 Superimposed on the image, the isophot map in arbitrary units
 shows the shape of a deformed galaxy.
 \label{fig3}}
\end{figure}

We constructed three models for the gravitational lens system.
The first model, which is unconstrained, consists of a
\textit{single }cluster and data from arcs A and B to make the
fit. Our best fit for the single component model predicts a
third image which is slightly demagnified (m$_{F814W}\approx
23$) in a position where no object is detected in the HST
exposures. We explored the field near its predicted position and
identified an object around $6\arcsec$ to the north
($\rm{R.A.}\,20\degr\,56\arcmin\,22 \farcs 5 \ \rm{and} \
\rm{DEC.}\,-04^h\,37^m\,37^s$) which seems to be a deformed
image of a galaxy. Fig.~\ref{fig3} shows the isocontour map
superimposed onto the optical image of this blurry object. We
measured the magnitude of this counter-image candidate
(m$_{F814W}=23.37\pm0.05$), and we found that it is similar to
the magnitude predicted by our model. Moreover, its F702W-F814W
color turns out to be analogous to arc A, $0.46 \pm 0.07$, (see
discussion below).

\begin{figure}
 \epsscale{.80}
 \plotone{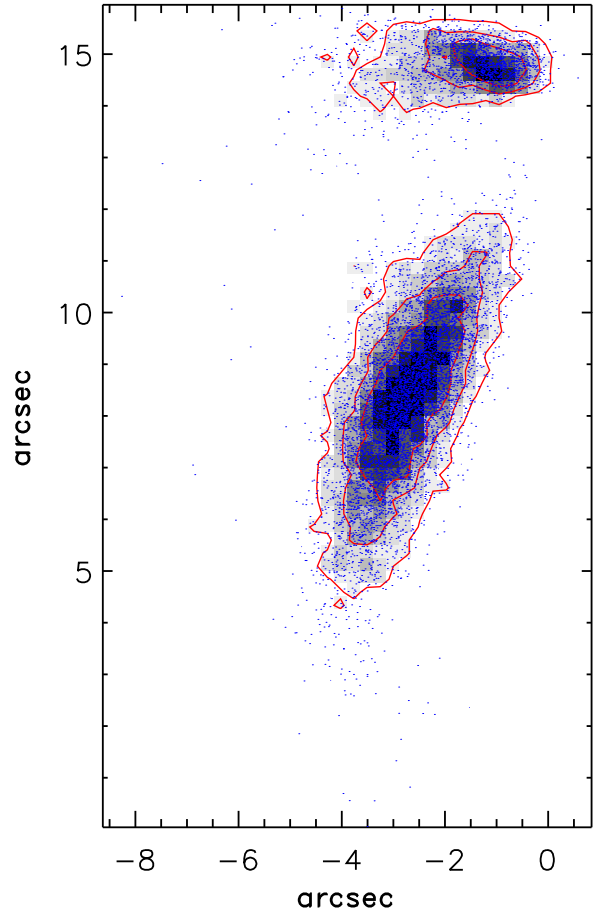}
 \caption{Two-dimensional distribution of the position of the
 cluster center for the SINGLE ABC model. Two clumps are easily
 distinguished. We used these clumps to construct the DOUBLE
 ABC model. The points show the results from the different trials
 of the MCMC optimization.\label{fig4}
 }
\end{figure}

In the second model, we repeated the previous fit but including
the third image candidate. With this third image the model is
well constrained because we have 8 constrictions and only 6
parameters. The fit was very accurate for the position but the
total $\chi^{2}$ is affected by the shape. It is interesting to
notice that in this well constrained model, there are two
different solutions in the XY-space. In Figure~\ref{fig4} we
show the XY parameter space for this lens model constructed
from all the trials coming out from the MCMC optimization
procedure. We think that this result is due to the fact that
the lens is a double cluster;  we used the regions defined by
these clumps to construct a third model, the \textit{double}
component model, that accounts for a bimodal distribution of
mass.

The results of these fits are summarized in Table~\ref{tbl-3}.
Column 1 identifies the model: \textsc{single} for models with
one component and \textsc{double }for the model with two
components; AB and ABC indicate that the model consists of arcs
A and B, or these two arcs plus the third image C. Columns 2 and
3 show the position in arcseconds relative to the BCG. Columns
4-8 list the parameters associated to NFW profile, ellipticity,
position angle, concentration, scale radius and $\sigma_s$,
respectively. Column 9 and Column 10 show the
$\chi^{2}$ for the image positions only and for the whole fit,
respectively.

\subsection{Mass profile}

\begin{figure}
 \epsscale{.80}
 \plotone{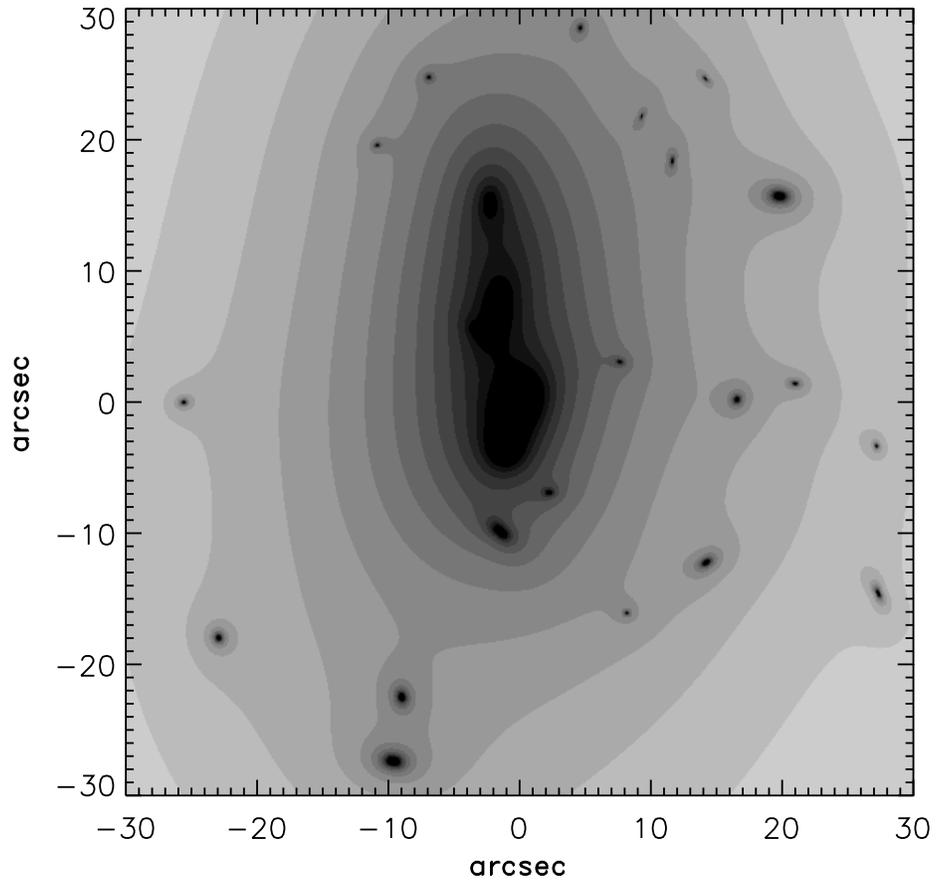}
 \caption{Projected mass distribution derived from the double
 component lens model. Note the elongation in the cluster
 mass which traces the distribution of the galaxies.
  \label{fig5}
 }
\end{figure}

\begin{figure}
 \epsscale{.80}
 \plotone{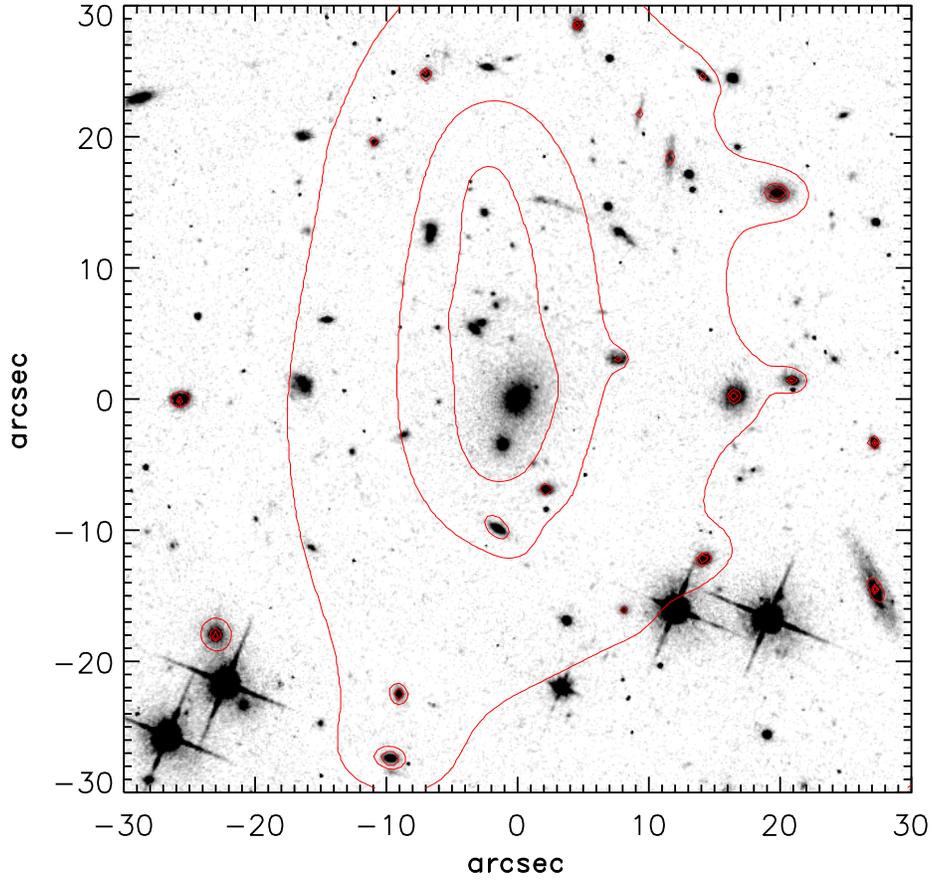}
 \caption{
HST WFPC2 F814W image of MS2053. The contours correspond to
the projected surface densities of $ 2.3, 4.8, 7.3 \times10^{10}$
M$_\odot$ arcsec$^{-2}$. As well as in Figure~\ref{fig5} we can
appreciate that the shape of the main component is elongated.
 \label{fig6} }
\end{figure}

For the purpose of quantifying the mass distribution, we traced
a projected mass map from the best-fit model, Double ABC
(Figure~\ref{fig5}). Figure~\ref{fig6} shows the corresponding
isocontour map superimposed onto the F814W image of MS\,2053.
By integrating this two dimensional map, we can get the total
mass of the cluster. In other words, we can determine the mass,
making an azimuthal average of its mass in circles around the
BCG.

\begin{figure}
\epsscale{.80}
\plotone{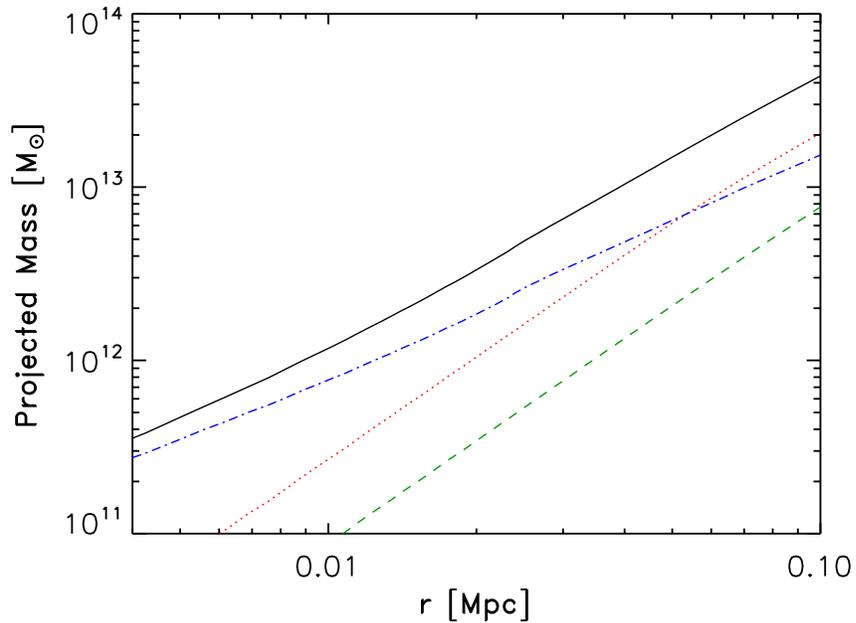}
\caption{The projected mass as a function of the aperture radius
 measured from the BCG for the three components of the
 cluster. The red-doted line and the green-dashed line
 illustrates the mass profile for
 \emph{Component 1} and  \emph{Component 2},
 respectively. The blue-dot-dashed line shows the
 masses of the individual cluster components.
 The black-solid line is the total mass of the cluster.
 \label{fig7}
}
\end{figure}

In the figure~\ref{fig7} we compare the projected mass as a
function of radius for the three components of the cluster: The
two clusters (\emph{Component 1} and \emph{Component 2}) and
the individual galaxies. To calculate the contribution of the
individual galaxies, we extrapolate the masses of the
individual cluster components to their virial radii and add
them. We can appreciate that within the central 0.8\,Mpc, the
dominant component corresponds to the mass of the galaxies.
However, at the arc radius (r $\sim 0.1$\,Mpc) the
\emph{Component 1} has the main contribution to the lens mass.
Figure ~\ref{fig7} also show the total mass of the cluster as
the sum of these three components.

Figure~\ref{fig8} shows the sum of the projected mass of
\emph{Component 1} and  \emph{Component 2} (without including
the galaxies) as a function of radius. We have extrapolated the
NFW profiles up to 1 Mpc in order to compare with the virial
mass estimate by \citet{tra05}, and with a $\beta$ model
derived by \citet{vik02}. The vertical line shows the distance
of arc A with respect to the BCG, and it indicates the limit
below which the mass estimate from strong lensing is reliable.
The mass determined using X-ray data depends on assumptions
involving spherical symmetry and hydrostatic equilibrium.
However, MS\,2053 shows evidence of being two clusters enduring
a merger process \citep{tra05}, and therefore it is not
expected to find a good agreement between the $\beta$ model and
the strong lensing mass profile (see \S\,\ref{discuss}).

\section{DISCUSSION}\label{discuss}

We found that the colors for the ABC arc system turned out to
be virtually the same, with arc B slightly bluer but still
coinciding with the color of arc A and image C within the
errors. This result supports the assumption that images A, B
and C belong to the same source. Despite this fact a serious
concern, is the lack of specularity; arcs A and B should be
mirror images of each other, for which there is no evidence in
the HST image. However, this would not be the first case of
lack of specularity. For example, the cluster A\,383 shows a
giant tangential arc with substructures and with no evidence of
being formed by two mirror images; these substructures have
different colors \citep{smi01} and lie at the same redshift, $z
= 1.01$ \citep{san04}. \citet{smi05} have shown that these arc
features can be understood in terms of just one galaxy
positioned over the tangential and radial caustic lines
simultaneously, with three different portions of the same
galaxy producing the arc substructure. We think that something
similar may be occurring in MS\,2053, that is the proximity of
the source to the caustic line (see Figure~\ref{fig9}). The
color distribution in galaxies often changes at different
optical radii (e.g., bulge, disk and halo populations). In the
case of a spiral galaxy for which the disk (but not the bulge)
is located over a caustic line, part of the disk is mapped onto
a stretched arc such as arc B. This effect may explain the
slight difference, if any, for the arc B color, the difference
in shape between arcs A and B, and the lack of specularity
between the arcs.

For the rest of this section we will assume that the three
images A, B, and C correspond to the same source. This
assumption, along with the evidence presented in the following
lines, in the sense that the quality of the fits is better for
the model with two cluster components, justifies that our
discussion shall be focused on our model 3. The $\chi^{2}$ for
the whole fit is $< 2.0$ for the \textsc{double\,abc} model and
$> 3.0$ for the other two, favoring the bimodal model (cf.
$\chi_{tot}^{2}$ values in Table~\ref{tbl-3}). In addition the
$\sigma_s$ values, $746 \pm 114$, and $392 \pm 133$ for
\emph{Component\,1} and \emph{Component\,2} respectively, agree
within the errors with the values reported by \citet{tra05};
but the other models yield larger velocities (\textit{c.f.}
Tables~\ref{tbl-1} and \ref{tbl-3}).

\citet{hoe02} suggested that the mass distribution in MS\,2053
is elongated in the direction of the giant arc; this projected
spatial distribution of the cluster was confirmed by
\citet{tra05} and it was attributed to a real elongation in the
spreading of cluster members. The critical lines for our model,
shown in Figure~\ref{fig9}, reveal that the main component of
the double cluster has a position angle of $\approx
80^{\circ}$, and supports the existence of a mass
elongation. In addition, the mass map as well as the isocontour
map of the projected mass distribution (see Figure~\ref{fig5}
and Figure~\ref{fig6}) reflect MS\,2053's elongated
distribution. The offset in the position of the cluster
components with respect to the BCG (see Table~\ref{tbl-3}) also
confirms the cluster asymmetry. Similar offsets have been found
by other authors for different clusters
\citep[e.g.][]{ogu04,cov05}, and the effects can reach values
up to 120 kpc in unrelaxed clusters  \citep{smi05}.

Several researches have reported large concentration values by
fitting NFW profiles to clusters of galaxies
\citep[e.g.][]{gav03, kne03,bro05}. However, these results were
obtained assuming spherically symmetric mass distributions,
although the CDM model predicts triaxial halos as consequence
of the collisionless dark matter \citep{jin02}. Using N-body
simulations, \citet{clo04} argue that the halo triaxiality
affects the concentration parameter measurements obtained
through gravitational lensing. \citet{gav05} concludes that a
prolate halo aligned toward the line of sight is a natural
explanation for the high concentration found in MS 2137-2353;
but he also leaves open the possibility that the lensing mass
estimates can be subestimated. \citet{ogu05} also investigated
the importance of the halo triaxiality and demonstrated that
this could cause a significant bias in estimating the virial
mass and concentration parameter from the lensing information.

In contrast, the concentration parameters obtained either for
the single or for the double component clusters may be slightly
lower (\textit{c.f.}Tables~\ref{tbl-1} and \ref{tbl-3}),
although they agree, within the errors, with the expected
values . Following \citet{gav05}, a halo with axis ratio
$\geqslant$ 1 could explain the possible small discrepancies
between our model and those inferred by \citet{bul01} and
\citet{eke01} ($c \sim 4$ and $c \sim 5$) for a cluster with
$1.1\times10^{15}$ M$_{\sun}$ and $1.1\times10^{14}$ M$_{\sun}$
in a $\Omega_{\Lambda}$-dominated universe.

\begin{figure}
 \epsscale{.80}
 \plotone{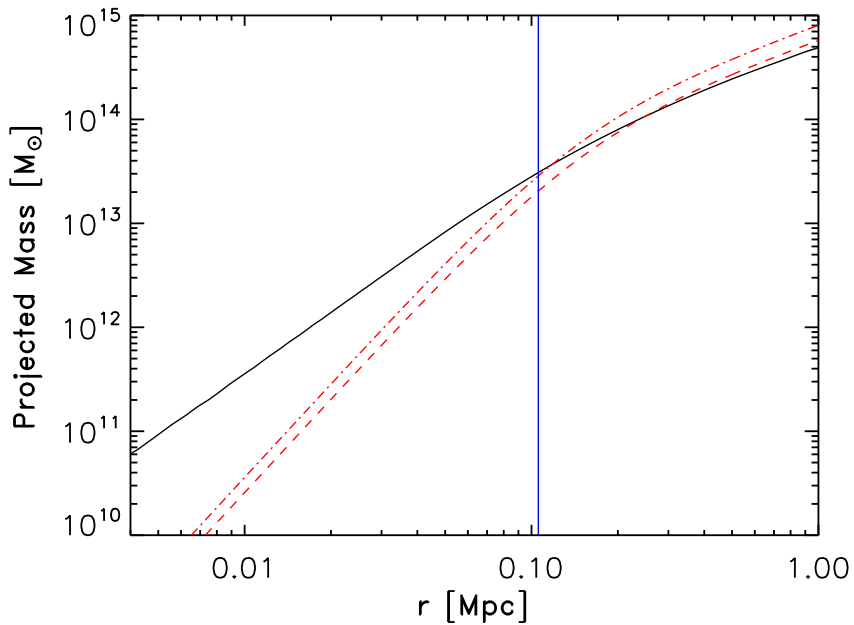}
 \caption{The projected mass for the NFW profile
 as a function of the aperture radius
 measured from the BCG.
 The black-solid line shows the mass profile derived from the
 sum of  \emph{Component 1} and  \emph{Component 2}
 in the double component model. The vertical blue line indicates the
 radius of the AB arc system. The mass estimate inside this radius
 is reliable. The mass outside the AB arc system has been
 extrapolated from the NFW model. The dashed line illustrates a
 $\beta$ model derived from data of \citet{vik02}. The
 dot-dashed line shows the same $\beta$ model corrected for the
 lack of hydrodynamical equilibrium in a merging cluster
 \citep{ras06}.\label{fig8}
 }
\end{figure}

\begin{figure}
 \epsscale{.80}
 \plotone{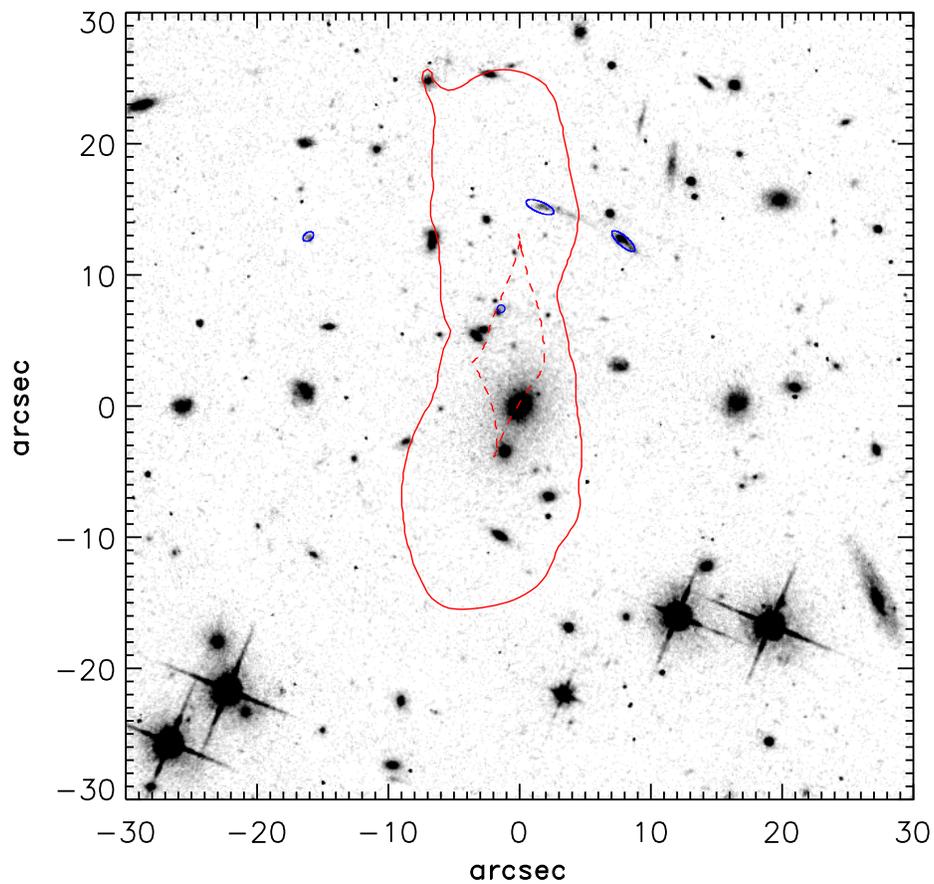}
 \caption{The external critic (continuous) and caustic (dashed)
 lines for the DOUBLE ABC model and the position of the lensed
 galaxy (shown as a small circle) in the source plane.
 The ellipses are the images predicted by our best
 model.\label{fig9}}
\end{figure}

A lower concentration changes the mass estimate. At large
radii,  the effect tends to vanish and the weak lensing and
true masses are the same  (see Figure~\ref{fig8} ), but inside
the arcs (r $\lesssim 0.1$\,Mpc) the effect is not negligible
producing an overestimate with respect to its actual X-ray
mass. But the differences in the masses can also be explained
as a result of the merging process. \citet{ras06} studied a set
of five galaxy clusters resolved at high resolution in a
hydrodynamic simulation, examining the systematics affecting
the X-ray mass estimates. They showed that for a cluster
undergoing a merger, the assumption of hydro-dynamical
equilibrium led to the underestimation of mass by 20$\%$, and
that a $\beta$ model gave even more discrepant results with
typical deviations of about 40$\%$. In Figure~\ref{fig8} we
compare the strong lensing mass profile to the X-ray mass
profile corrected for the 40$\%$ underestimation; we can see
that with this correction, the X-ray mass estimate agrees with
the mass obtained from the NFW components at the positions of
the arcs, but for smaller radii the X-ray mass is significantly
lower. It is important to notice that we did not have included
the masses of the individual cluster components, and thus we
are not considering in Figure~\ref{fig8} the total mass within
the arcs. In addition, the extrapolation of the NFW mass
inferred from gravitational lensing up to 1\,Mpc from the BCG,
well beyond the distance of the arcs, is a factor $\sim 2$
lower than the mass shown in Table~\ref{tbl-1} obtained by
\citet {tra05}.

NFW ellipsoids have been successfully used for strong lensing
modelling in galaxy clusters \citep[e.g.][]{com06}. However,
the non-linearity of the gravitational lensing effect, as well
as its high dependence on the core densities, the asymmetries
of the mass distribution, and the cluster neighborhoods, put
limits to the use of analytical solutions in the case of non
relaxed systems, such as MS~2053. In this sense, the work of
\citet{tor04}, who have computed the strong lensing effect in a
merging cluster with complex and irregular mass distribution,
have shown that the lensing cross-sections can grow by one
order of magnitude and that during the merger, the shape of the
critical and caustics lines changes substantially. These topics
have been discussed in a recent paper by \citet{fed06}, who
propose a fast method to calculate cross sections for complex
and asymmetric mass distributions.

\section{CONCLUSION}\label{conclus}

We have modeled the mass distribution of the cluster MS\,2053
using the LENSTOOL ray-tracing code developed by \citet{kne93}.
The fits were performed considering both a single and a double
NFW profiles, and 23 Galaxy-scale mass components as
perturbations to the cluster potential. We used ESISs for the
individual galaxies and set constraints to their parameters
using observational data. We measured the arc positions and
shapes, and the galaxies positions using the HST archive image,
and calculated the NFW parameters for the single and double
galaxy cluster models. Our main results can be summarized as
follows:
\begin{enumerate}
    \item The XY parameter space of the \textsc{single ABC}
        model has a bimodal distribution that strongly
        suggests a double cluster undergoing a merger
        process. The quality of the fits also favor a model
        with a bimodal mass distribution
    \item The models consistently predict a third slightly
        demagnified counter-image for the AB arc system.
    \item We found a candidate for this counter-image near
        the predicted position. This candidate shows a
        fuzzy object that can be the image of a distorted
        galaxy due to the effect of gravitational lensing. Besides,
        it has the same F702W-F814W colors as arcs A and B.
    \item Using the strong lensing effect we confirm the
        asymmetry and the elongation in the mass
        distribution of MS\,2053 reported by other authors
        \citep{hoe02,tra05}, and we estimate the total lensing mass
        within the arcs to be $4.7\times10^{13}$ M$_\odot$.
        We find a significant discrepancy between the mass
        estimates from lensing and X-ray measurements, that
        we attribute to the fact that the cluster is not in
        hydro-dynamical equilibrium.
    \item The concentration parameters obtained either for
        the single or for the double component clusters may
        be slightly lower than the expected values. These
        possible small discrepancies can be explained by
        asymmetries in the mass distribution and the
        projection effects.

  \end{enumerate}

Much more work is necessary in order to understand MS\,2053
mass distribution.  Future spectroscopic follow-up of arcs B
and C will provide a test of our model. Also, an extensive
study of the internal velocity dispersion of the cluster
members can be used to further improve the accuracy of the mass
reconstruction.

\acknowledgments

This work was partially supported by the DGAPA-UNAM grant
IN113002. T. Verdugo also acknowledges the scholarship support
by CONACyT (Register Number 176538), and CONACyT grant 54799.

Some of the data presented in this paper were obtained from the
Multimission Archive at the Space Telescope Science Institute
(MAST). STScI is operated by the Association of Universities
for Research in Astronomy, Inc., under NASA contract
NAS5-26555. Support for MAST for non-HST data is provided by
the NASA Office of Space Science via grant NAG5-7584 and by
other grants and contracts". The Dark Cosmology Centre is
funded by the Danish National Research Foundation.

The author thank Vladimir Avila for useful comments, and the
anonymous referee for invaluable remarks and suggestions. We
also thank J. Benda for helping with proofreading.

\appendix

\section{The Einstein angle for a NFW profile}

Consider a spherical NFW density profile acting like a lens.
The analytical solutions for this lens were given by
\citet{bar96} and have been studied by different authors
\citep{wri00,gol02,men03}. The positions of the source and the
image are related through the equation:

\begin{equation}\label{eq:u}
\vec u^{S} = \vec u^{I} - \nabla\varphi( \vec u^{I}) =  \vec u^{I} -  \vec \alpha( \vec u^{I})
\end{equation}

\noindent where $\vec u^{I}$ and $\vec u^{S}$ are the angular
position in the image and in the source planes, respectively.
$\vec \alpha$ is the deflection angle between the image and the
source and $\varphi$ is the two-dimensional lens potential. We
introduce the dimensionless radial coordinate $\vec x=\vec
u^{I}/u_s^{I}$, where $u_s^{I}=r_s/D_{OL}$, and $D_{OL}$ the
angular diameter distance between the observer and the lens. In
the case of an axially symmetric lens, the relations become
simpler, as the position vector can be replaced by its norm.

The deflection angle then becomes \citep{gol02}:

\begin{equation}\label{eq:alpha}
\vec \alpha(x) = 4k_s\frac{u^{I}}{x^{2}}g(x)\hat{e_x}
\end{equation}

\noindent Where $g(x)$ is a function related with the surface
density inside the dimensionless radius $x$, and is given by
\citep{bar96}:

\begin{equation}
g(x) = \left\{ \begin{array}{ll}
\ln\frac{x}{2} + \frac{1}{\sqrt{1-x^{2}}} \, \rm{arccosh}\frac{1}{x} & \textrm{if $x<1$}\\

1+\ln\frac{1}{2} & \textrm{if $x = 1$}\\

\ln\frac{x}{2} + \frac{1}{\sqrt{x^{2}-1}}\arccos\frac{1}{x} & \textrm{if $x>1$}\\
\end{array} \right.
\end{equation}

\noindent Where
$k_s=r_s\rho_{crit}\delta_{c}\Sigma_{crit}^{-1}$, the lensing
strength, is an estimate of the convergence parameter
\citep[see the discussion in][]{ogu04}. The quantity
$\Sigma_{crit}=(c^{2}/4\pi G)(D_{OS}/D_{OL}D_{LS})$ is the
critical surface mass density for lensing, and $D_{ LS }$ and
$D_{ OS }$ are the angular diameter distances between the lens
and the source, and the observer and the source, respectively.

If we calculate the deflection angle at radius $r_s$, we obtain
$x=1$, and equation \ref{eq:alpha} can be expressed as:

\begin{equation}
\vec \alpha(x) =    \frac{16\pi G\rho_{crit}\delta_{c}r_s^{2}D_{LS}}{c^{2}D_{OS}}\left(1+\ln{\frac{1}{2}}\right) \hat{e_x}
\end{equation}

\noindent Hence, the lens equation \ref{eq:u} can be written
as:

\begin{equation}
\vec u^{S} = \vec u^{I} \left(1+{\frac{u_{E,r_s}^{I}}{u^{I} }}\right)
\end{equation}

\noindent Where $u_{E,r_s}^{I}$ is the Einstein angle that for
the case $x=1$ is expressed by:

\begin{equation}\label{eq:ue}
u_{E,r_s}^{I}  =    \frac{16\pi G\rho_{crit}\delta_{c}r_s^{2}D_{LS}}{c^{2}D_{OS}}\left(1+\ln{\frac{1}{2}}\right)
\end{equation}

\noindent For a circularly symmetric lens the Einstein radius
is given by $u_{E}^{I} =4\pi\sigma^{2}D_{LS}/c^{2}D_{OS}$.
Therefore, equation \ref{eq:ue} takes the form:

\begin{equation}
u_{E,r_s}^{I}  =    \frac{4\pi\sigma_s^{2} D_{LS}}{c^{2}D_{OS}}
\end{equation}

\noindent Where  $\sigma_s^{2} = 4 \left (1+\ln{\frac{1}{2}}
\right)G r_s^{2} \rho_{crit}\delta_{c}$. Therefore, $\sigma_s$
represents the velocity at radius $r_s$. We adopt this
characteristic velocity in order to compare the velocity
predicted by our models with the velocity dispersion measured
in dynamical studies.

\clearpage

\clearpage

\begin{deluxetable}{lcccccc}
\tabletypesize{\scriptsize}

\tablecaption{Cluster parameters\label{tbl-1}}
\tablewidth{0pt}

\tablehead{ \colhead{Group}   &  \colhead{$z$} &
\multicolumn{1}{c}{$M_{vir}$} &  \multicolumn{1}{c}{$\sigma$}
&    \multicolumn{1}{c}{$\sigma\tablenotemark{a}$} &
 \multicolumn{1}{c}{$c\tablenotemark{b}$} &  \colhead{Reference}\\

 \colhead{} & \colhead{} & \colhead{$(M_{\odot})$} &
\colhead{(km s$^{-1}$)} & \colhead{(km s$^{-1})$} & \colhead{} & \colhead{}}

\startdata

$MS\,2053$ & $0.5866 \pm 0.0011$ & $3.7\times10^{15} $ & $1523 \pm 95$ & $886_{-139}^{+121}$ & $2.9$ &  $1,2$ \\
$MS\,2053-A$ & $0.5840 \pm 0.0005$ & $1.1\times10^{15} $ & $\phantom{0}865 \pm 71$ & $$  & $3.8$ & $1$ \\
$MS\,2053-B$ & $0.5982 \pm 0.0003$ & $1.1\times10^{14} $ & $\phantom{0}282 \pm 51$ & $$ & $5.4$  & $1$\\

\enddata
\tablenotetext{a}{ Weak lensing data}
\tablenotetext{b}{Calculated from \citet{eke01} }
\tablerefs{
(1) \citet{tra05}; (2) \citet{hoe02}}
\end{deluxetable}

\clearpage

\begin{deluxetable}{cccccrccc}
\tabletypesize{\scriptsize}

\tablecaption{Data of the galaxies\label{tbl-2}}
\tablewidth{0pt}

\tablehead{ \colhead{ID Name}   &  \multicolumn{1}{c}{$\alpha$}
&  \multicolumn{1}{c}{$\delta$} &  \multicolumn{1}{c}{$F814W$} &  \colhead{$\epsilon$\tablenotemark{b}}
&    \multicolumn{1}{c}{$\theta_0$}
&  \colhead{z} &  \multicolumn{1}{c}{$\sigma_0$}  &  \colhead{Reference}  \\

 \colhead{} & \colhead{(J2000)} & \colhead{(J2000)} & \colhead{(mag)} &
 \colhead{ } & \colhead{$(^{\circ})$} & \colhead{} &
\colhead{(km s$^{-1})$} &  \colhead{}
}

\startdata

 $a$ & $20:56:21.4$ & $-04:37:50$ & $19.13\pm0.01  $ & $0.259$ & $\phantom{0}63.9$ & $0.583$ & $292.0\pm10.0  $ & $1$\\
$b$ & $20:56:21.5$ & $-04:37:54$ & $20.24\pm0.01  $& $0.122$ & $\phantom{0}81.8$ & $0.5894$ & $234.0\pm23.0 $ & $1$\\
$c$ & $20:56:21.5$ & $-04:38:00$ & $20.93\pm0.02  $& $0.577$ & $-41.1$ & $0.5724$ & $151.0\pm17.0 $& $1$\\
$d$ & $20:56:20.1$ & $-04:37:35$ & $20.29\pm0.01  $& $0.368$ & $-\phantom{0}7.5$ & $0.5846$ & $169.0\pm14.0$ & $1$\\
$e$ & $20:56:20.5$ & $-04:38:02$ & $20.84\pm0.01  $& $0.473$ & $\phantom{0}32.9$ & $0.5737$ & $129.0\pm15.0 $ & $1$\\
$f$ & $20:56:22.0$ & $-04:38:13$ &  $21.47\pm0.02  $& $0.278$ & $\phantom{0}80.2$ & $0.5800$ & $138.0\pm13.0 $ & $1$\\
$g$ & $20:56:23.0$ & $-04:38:08$ & $20.19\pm0.01  $& $0.136$ & $-74.2$ & $0.5854$ & $124.0\pm13.0 $ & $1$\\
$h$ & $20:56:22.1$ & $-04:38:18$ & $21.12\pm0.02  $& $0.374$ & $-\phantom{0}9.8$ & $0.5835$ & $181.0\pm13.0 $ & $1$\\
$i$ & $20:56:21.3$ & $-04:37:57$ &  $21.09\pm0.02  $ & $0.299$ & $\phantom{00}7.1$ & $0.5854$ & $<200.0$ & $2$\\
$k$ & $20:56:19.6$ & $-04:38:05$&  $19.98\pm0.01  $ & $0.867$ & $-66.5$ & $0.5912$ & $<200.0$ & $2$\\
$l$ & $20:56:19.6$ & $-04:37:54$&   $21.38\pm0.02  $ & $0.270$ & $-68.3$ & $0.5824$ & $<200.0$& $2$\\
$m$ & $20:56:20.0$ & $-04:37:49$& $20.88\pm0.01  $ & $0.361$ & $-10.1$ & $0.5813$ & $<100.0$ & $2$\\
$n$ & $20:56:21.1$ & $-04:37:22$&  $21.34\pm0.02  $ & $0.240$ & $\phantom{0}76.1$ & $0.5764$ & $\phantom{<1}86.0$ & $3$\\
$o$ & $20:56:21.9$ & $-04:37:25$&  $21.67\pm0.03  $ & $0.063$ & $\phantom{0}26.3$ & $0.5799$ & $<200.0$ & $2$\\
$p\tablenotemark{a}$ & $20:56:23.2$ & $-04:37:50$& $20.83\pm0.01  $ & $0.203$ & $\phantom{0}11.7$ & $0.5990$ & $<100.0$ & $2$\\
$q\tablenotemark{a}$ & $20:56:20.9$ & $-04:37:47$& $21.74\pm0.03  $ & $0.335$ & $-\phantom{0}9.0$ & $0.5978$ & $<100.0$ & $2$\\
$r\tablenotemark{a}$ & $20:56:20.3$ & $-04:37:50$&  $20.04\pm0.01  $ & $0.116$ & $\phantom{0}65.5$ & $0.5990$ & $<100.0$ & $2$\\
$s\tablenotemark{a}$ & $20:56:20.7$ & $-04:37:32$& $21.41\pm0.02  $ & $0.855$ & $\phantom{0}84.8$ & $0.5993$ & $<100.0$ & $2$\\
$t$ & $20:56:20.9$ & $-04:38:06$&  $21.90\pm0.03  $ & $0.093$ & $-\phantom{0}4.4$ & $0.5827$ & \nodata & $$\\
$u$ & $20:56:21.7$ & $-04:37:45$& $21.04\pm0.02  $ & $0.546$ & $-47.6$ & $0.5876$ & \nodata & $$\\
$v$ & $20:56:22.2$ & $-04:37:31$& $21.99\pm0.04  $ & $0.217$ & $\phantom{0}22.3$ & $0.5892$ & \nodata & $$\\
$w$ & $20:56:20.8$ & $-04:37:28$& $22.57\pm0.06  $ & $0.910$ & $\phantom{0}68.7$ & $0.5939$ & \nodata & $$\\
$x$ & $20:56:20.5$ & $-04:37:26$ & $21.90\pm0.04  $ & $0.798$ & $-48.9$ & $0.5782$ & \nodata & $$\\

\enddata

\tablenotetext{a}{ Galaxies that belong to the second cluster
component, according to \citet{tra05} } \tablenotetext{b}{ The
ellipticities are defined as $\epsilon = \frac{ a^{2}-b^{2} }{
a^{2}+b^{2} } $, where $a$ and $b$ are respectively
the semi-major and semi-minor axis of the
elliptical shape.  }

\tablerefs{ (1) Wuyts et
al. 2004; (2) Tran et al. 2005; (3) Tran et al. 2003 .}

\end{deluxetable}

\clearpage

\begin{deluxetable}{lrrrrrrrrr}
\tabletypesize{\scriptsize}
\rotate
\tablecaption{Best-fitting parameters\label{tbl-3}}
\tablewidth{0pt}

\tablehead{ \colhead{Model}    &  \multicolumn{1}{c}{X} &
\multicolumn{1}{c}{Y}  &  \colhead{$\epsilon$} &
\multicolumn{1}{c}{$\theta$} & \colhead{c} &
\multicolumn{1}{c}{$r_s$} & \multicolumn{1}{c}{$\sigma_s$}
& \colhead{$\chi_{pos}^{2}$} & \colhead{$\chi_{tot}^{2}$}\\

\colhead{} & \colhead{$('')$} & \colhead{$('')$} & \colhead{ }
& \colhead{$(^{\circ})$} & \colhead{} & \colhead{(kpc)} &
\colhead{(km s$^{-1})$} & \colhead{} & \colhead{}}

\startdata


SINGLE AB       & $-4.1 \pm 3.3$    & $\phantom{1}6.6 \pm 4.8$  & $0.24 \pm 0.15$   & $104 \pm 37$                          & $2.6 \pm 0.7$ & $1259 \pm 312$                        & $ 1251 \pm 165$           & $1.3\phantom{0}$  & $4.6$\phantom{0}\\
SINGLE ABC      & $-2.5 \pm 0.9$    & $\phantom{1}9.7 \pm 3.0$  & $0.25 \pm 0.11$   &  $\phantom{1}83 \pm \phantom{1}4$     & $2.4 \pm 0.6$ & $1265 \pm 290$                        & $1136 \pm 100$            & $0.2\phantom{0}$  & $3.4$\phantom{0}\\
DOUBLE ABC      & $$                & $$                        & $$                & $$                                    & $$            & $$                                    & $$                        & $1.2\phantom{0}$  & $1.7$\phantom{0}\\
~~Component 1   & $-1.7 \pm 1.4$    & $\phantom{1}7.2 \pm 2.5$  & $0.34 \pm 0.16$   & $\phantom{1}81 \pm 18$                & $3.1 \pm 0.9$ & $\phantom{1}634 \pm 159$              & $\phantom{1}746 \pm 114$  & $$                & $$\\
~~Component 2   & $-2.3 \pm 1.9$    & $15.1 \pm 1.1$            & $0.27 \pm 0.19$   & $\phantom{1}89 \pm 41$                & $4.9 \pm 2.1$ & $\phantom{1}202 \pm \phantom{1}50$    & $\phantom{1}392 \pm 133$  & $$                & $$\\
\enddata

\end{deluxetable}


\end{document}